\documentclass[apj]{emulateapj}
\usepackage{apjfonts}
\usepackage{amsbsy}

\def\hmpc{$h^{-1}$Mpc}
\def\hkpc{$h^{-1}$kpc}

\def\hmsol{$h^{-1}$M$_\odot$}
\def\kms{km\,s$^{-1}$}

\def\nsat{\langle N_{\rm sat}\rangle_M}
\def\ncen{\langle N_{\rm cen}\rangle_M}
\def\mmin{M_{\rm min}}

\def\asat{\alpha_{\rm sat}}

\def\om{\Omega_m}
\def\omb{\Omega_b}
\def\s8{\sigma_8}

\def\lcdm{$\Lambda$CDM}

\def\x2{$\chi^2$}
\def\hmsol{$h^{-1}\,$M$_\odot$}

\def\ngavg{\bar{n}_g}

\def\NNm1{\langle N(N-1) \rangle}

\def\slogm{\sigma_{{\rm log}M}}
\def\m_star{M_\ast}

\def\rhob{\tilde{\rho}}

\def\lcdm{$\Lambda$CDM}
\def\slogm{\sigma_{{\rm log}M}}
\def\om{\Omega_m}

\def\omb{\Omega_b}
\def\s8{\sigma_8}

\def\hmpc{$h^{-1}\,$Mpc}
\def\hkpc{$h^{-1}\,$kpc}

\def\x2{$\chi^2$}
\def\hmsol{$h^{-1}\,$M$_\odot$}

\def\kms{km\,s$^{-1}$}

\def\mmin{M_{\rm min}}

\def\sigmaM{\sigma_{\log M}}

\def\nsat{\langle N_{\mbox{\scriptsize sat}}\rangle_M}

\def\ncen{\langle N_{\mbox{\scriptsize cen}}\rangle_M}

\def\ngavg{\bar{n}_g}

\def\NNm1{\langle N(N-1) \rangle}

\def\sigmaM{\sigma_{\log M}}

\def\p0{P_0(r)}

\def\rnb{R_{\rm nb}}
\def\lg{{\it l}_g}
\def\lstar{L_\ast}
\def\rv{R_{\rm void}}
\def\rhob{\bar{\rho}}

\begin{document}

\title{The Void Phenomenon Explained}

%\author{ Authors } 
\author{ Jeremy L. Tinker$^{1}$ \& Charlie Conroy$^{2}$}
\affil{$^{1}$Kavli Institute for Cosmological Physics \& Department of
  Astronomy and Astrophysics, University of Chicago\\ $^{2}$Department
  of Astrophysical Sciences, Princeton University, Princeton, NJ
  08544, USA}

\begin{abstract}

  We use high-resolution N-body simulations, combined with a halo
  occupation model of galaxy bias, to investigate voids in the galaxy
  distribution. Our goal is to address the ``void phenomenon'' of
  \cite{peebles:01}, which presents the observed dearth of faint
  galaxies in voids as a challenge to the current cosmology. In our
  model, galaxy luminosity is determined only as a function of dark
  matter halo mass. With this simple assumption, we demonstrate that
  large, empty voids of $\sim 15$ \hmpc\ in diameter are expected even
  for galaxies seven magnitudes fainter than $L_\ast$. The predictions
  of our model are in excellent agreement with several statistical
  measures; ($i$) the luminosity function of galaxies in underdense
  regions, ($ii$) nearest neighbor statistics of dwarf galaxies,
  ($iii$) the void probability function of faint galaxies. In the
  transition between filaments and voids in the dark matter, the halo
  mass function changes abruptly, causing the maximum galaxy
  luminosity to decrease by $\sim 5$ magnitudes over a range of $\sim
  1$ \hmpc. Thus the boundary between filaments and voids in the
  galaxy distribution is nearly as sharp for dwarfs as for $\sim
  \lstar$ objects. These results support a picture in which galaxy
  formation is driven predominantly by the mass of the host dark
  matter halo, and is nearly independent of the larger-scale halo
  environment. Further, they demonstrate that \lcdm, combined with a
  straightforward bias model, naturally predicts the existence of the
  void phenomenon.

\end{abstract}


%%%%%%%%%%%%%%%%%%%%%%%%%
\section{Introduction}
%%%%%%%%%%%%%%%%%%%%%%%%%

Galaxy redshift surveys have revealed a complex network of clusters,
filaments and walls. They have also demonstrated that the expansive
regions between these structures are nearly bereft of galaxies (see
\citealt{gregory_thompson:78, kirshner_etal:81, vogeley_etal:94} for
early work, and \citealt{hoyle_vogeley:04, croton_etal:04,
  conroy_etal:05, patiri_etal:06a, tinker_etal:07_voids} for results
from recent large-scale surveys). The surprisingly large size of
voids---up to $\sim 30$ \hmpc\ in diameter---and their apparent depth,
in terms of their luminosity density, have begged questions about
their formation mechanism. Why do faint galaxies avoid the voids
defined by their brighter brethren? Further, the few objects found in
and near underdense regions appear to represent a fair sample of the 
overall galaxy population.  Peebles (\citeyear{peebles:01}, hereafter
P01) presented these observations as ``the void phenomenon''. The
straightforward argument is that the deepest voids in the dark matter
distribution predicted by standard \lcdm\ are never completely empty;
they still contain mass and significant numbers of low-mass
halos. Therefore, voids should be the preferable environment of faint
dwarf galaxies. However, P01 demonstrates that dwarf galaxies tend to
be distributed similarly to brighter spirals, and the local void
exhibits a paucity of dwarfs. \cite{peebles:07} further shows that the
edge of the local void has the same sharp edge for both bright and
faint galaxies. If halos exist in voids, then where are the galaxies?

In this paper we present a quantitative model of galaxy bias that
allows us to directly address this question. We use the Halo
Occupation Distribution (HOD; e.g. \citealt{seljak:00, roman_etal:01,
  cooray_sheth:02, berlind_weinberg:02}) to specify the relationship
between galaxies and dark matter. In the HOD framework, this
relationship is quantified by $P(N|M)$, the probability that a halo of
mass $M$ contains $N$ galaxies of a given sample. Galaxy samples can
be defined by luminosity, color, star formation rate, or any other
galaxy property. For each sample, $P(N|M)$ will be different but can
be calibrated with observational data. The HOD has emerged as the
dominant tool for interpreting clustering measurements for a wide
range of redshifts and galaxy classes (see, for a tip of the iceberg,
\citealt{zehavi_etal:04, zehavi_etal:05, zheng_etal:07, vdb_etal:07,
  tinker_etal:07_pvd, tinker_etal:07_voids, chen:07, white_etal:07,
  padmanabhan_etal:08}). The results obtained from analysis of
observational data are in excellent agreement with the results of
semi-analytic models, hydrodynamic cosmological simulations, and
high-resolution collisionless simulations (\citealt{kravtsov_etal:04,
  zheng_etal:05, conroy_etal:06}). Our knowledge of the mapping
between galaxies and halos is now well-established and can be
confidently extended into regimes in which observational data are
lacking. Here we take HOD results calibrated on observed galaxy
samples that occupy $\gtrsim 10^{11}$ \hmsol\ halos and extrapolate to
lower masses in order to model the distribution of dwarf galaxies. We
will demonstrate that the predictions of this model are in excellent
agreement with the observations of P01: deep, wide voids are expected
within hierarchical structure formation, even for the faintest of
galaxies.

One of the difficulties in directly addressing the void phenomenon is
that the problem has not been clearly defined. The definition of a
dwarf galaxy varies in the literature, and the predictions of \lcdm\
in this regime have not been robustly specified. To elucidate the
problem, P01 compared the spatial distribution of ``ordinary''
galaxies to samples of ``test'' dwarf galaxies within the Optical
Redshift Survey (ORS; \citealt{santiago_etal:95}). The number density
of the ordinary sample is comparable to the number density of
$M_r<-16$ galaxies\footnote{For brevity, all galaxy magnitudes assume
  a Hubble constant $h\equiv H_0/100=1$. All $r$-band magnitudes have
  been $k$-corrected to $z=0$, as with the \cite{blanton_etal:05}
  luminosity function.} in the Sloan Digital Sky Survey (SDSS;
\citealt{blanton_etal:05}). Thus to give our model a fixed target we
stipulate that the void phenomenon begins at galaxies fainter than
$M_r=-16$, roughly Small Magellanic Cloud-type galaxies and fainter.

In \cite{tinker_etal:07_voids} we used the HOD to make predictions for
void probability statistics for galaxies brighter than $M_r=-19$. Our
predictions were in excellent agreement with our measurements from
SDSS Data Release 4 (\citealt{dr4}). Thus we concluded that there is
no conflict between the observed voids and those predicted by \lcdm\
for galaxies down to this magnitude (roughly $0.2\lstar$, given an
$M_\ast$ of $-20.44$ from \citealt{blanton_etal:03}). More importantly
for the model presented herein, the results of
\cite{tinker_etal:07_voids} support a model in which the luminosities
of galaxies are determined entirely by the mass of the host halo,
independent of the environment in which the halo formed. This allows
us to robustly predict the distribution of void galaxies within a
model that connects galaxies with halos through the halo mass only.

Several other studies have investigated galaxy voids and void galaxies
through a variety of techniques. \cite{mathis_white:02} and
\cite{benson_etal:03}, using semi-analytic models, concluded that
galaxies of $M_r\sim -18.5$ avoid the voids of the brighter
galaxies. These studies were limited by numerical resolution and were
not able to model the dwarf galaxies on which P01 based his
argument. \cite{furlanetto_piran:06} constructed an analytic framework
within which to explore the dependence of voids on galaxy luminosity,
building on earlier work modeling voids in the dark matter by
\cite{sheth_weygaert:04}.  However, no tests with numerical
simulations or comparison to observational results were made. The
model of \cite{furlanetto_piran:06} is an excellent tool for
understanding the trends seen in the data, but detailed analysis of
voids requires a simulation to produce the correct halo
distribution. As we will show, the change in the halo mass function
along the edge of a halo is of critical importance in the resulting
void distribution. Additionally, observational data are all measured
in redshift space, which is not incorporated into these analytic
models.

This paper is organized as follows: In \S 2 we present our HOD model
and the numerical simulations employed to create mock galaxy
distributions. In \S 3 we present the predicted void distribution,
making quantitative comparisons to several sets of observational data:
(a) the luminosity function of void galaxies, which has been measured
down to $M_r=-14$ by \cite{hoyle_etal:05}, (b) the nearest neighbor
statistics of P01, which probe galaxies fainter than $M_r=-16$, and
(c) the void probability function of galaxies down to $M_r=-17$, for
which we present new measurements from SDSS Data Release 6
(\citealt{dr6}). In \S 4 we discuss our results. In all calculations,
we assume a flat, \lcdm\ universe with $(\om, \s8, h, n,
\omb)=(0.3,0.9,0.7,1.0,0.04)$.

%%%%%%%%%%%%%%%%%%%%%%%%%%%
%        TABLE 1          %
%%%%%%%%%%%%%%%%%%%%%%%%%%%

\begin{deluxetable}{ccccccc}
\tablecolumns{7} 
\tablewidth{14pc} 
\tablecaption{Minimum Halo Mass as a Function of $r$-band magnitude threshold}
\tablehead{\colhead{$M_r$} & \colhead{$\ngavg$} & \colhead{$\mmin$}&\colhead{} & \colhead{$M_r$} & \colhead{$\ngavg$} & \colhead{$\mmin$}}
\startdata

-22 & $2.63 \times 10^{-5}$ & 14.16 & & -15 & $1.30 \times 10^{-1}$ & 10.60 \\
-21 & $1.23 \times 10^{-3}$ & 12.73 & & -14 & $2.12 \times 10^{-1}$ & 10.38 \\
-20 & $6.59 \times 10^{-3}$ & 11.97 & & -13 & $3.43 \times 10^{-1}$ & 10.16 \\
-19 & $1.58 \times 10^{-2}$ & 11.57 & & -12 & $5.57 \times 10^{-1}$ & 9.94  \\
-18 & $2.88 \times 10^{-2}$ & 11.29 & & -11 & $9.01 \times 10^{-1}$ & 9.72  \\
-17 & $4.83 \times 10^{-2}$ & 11.05 & & -10 & $1.46 \times 10^{-0}$ & 9.51 \\
-16 & $7.95 \times 10^{-2}$ & 10.82 & & & & \\

\enddata
\tablecomments{Galaxy densities are in units of $($\hmpc$)^{-3}$. All
  masses are in units of \hmsol. }
\end{deluxetable}

%%%%%%%%%%%%%%%%%%%%%%%%%
\section{Methods}
%%%%%%%%%%%%%%%%%%%%%%%%%

%%%%%%%%%%%%%%%%%%%%%%%%%%%%%%%%%%%%%%
\subsection{Halo Occupation Models}
%%%%%%%%%%%%%%%%%%%%%%%%%%%%%%%%%%%%%%

Halo occupation models generally contain two important mass
scales. The first is the mass at which halos become sufficiently large
to host a single galaxy brighter than a specified luminosity
threshold. The second is the mass at which halos become massive enough
to host additional satellite galaxies. Results from both theory and
observation have demonstrated that these masses scale in a
straightforward fashion with galaxy luminosity, allowing for confident
extrapolation to the dwarf galaxies on which we focus.

Functionally, the HOD breaks down the occupation of galaxies within
halos into two distinct components: central galaxies and satellite
galaxies. We use the standard parameterization for central galaxies
brighter than a given magnitude limit:

\begin{equation}
\label{e.ncen}
\ncen = \frac{1}{2}\left[ 1+\mbox{erf}\left(\frac{\log M - \log \mmin}{\sigmaM} \right)
	\right],
\end{equation}

\noindent where $\mmin$ is the mass at which a halo has a 50\%
probability of having a central galaxy above the defined luminosity
threshold, and $\slogm$ is the width of the transition between 0 and 1
galaxies, physically representing the scatter of mass at fixed
luminosity. The number of satellite galaxies scales as a power of the
host mass, with a cutoff scale set by the central occupation function,

\begin{equation}
\label{e.nsat}
\nsat = \ncen \times \left(\frac{M}{M_1}\right)^{\asat}
\end{equation}

\noindent where $M_1$ is the mass scale at which a halo has, on
average, one satellite brighter than the defined magnitude limit. The
inclusion of $\ncen$ in equation (\ref{e.nsat}) ensures that the
central galaxy is the brightest galaxy within a halo. The forms of
these equations are in good agreement with the results of
semi-analytic models of galaxy formation (\citealt{zheng_etal:05}),
high-resolution collisionless N-body simulations that resolve
substructure (\citealt{kravtsov_etal:04, conroy_etal:06}), hydrodynamic
simulations (\citealt{zheng_etal:05}), and analytic models of halo
substructure (\citealt{zentner_etal:05}).

It has been demonstrated that the parameters of the occupation
functions are nearly self-similar with luminosity, i.e., that
$\asat\approx 1$ and $M_1/\mmin \approx 20$. These results have been
found both in the theoretical results listed above and analyses of
clustering measurements from numerous surveys at multiple redshifts
(\citealt{zehavi_etal:05, zheng_etal:07, tinker_etal:07_pvd,
  vdb_etal:07}).  High-resolution N-body simulations have demonstrated
that the subhalo mass function is nearly self-similar with parent halo
mass, in good agreement with the HOD results obtained from
observations (\citealt{kravtsov_etal:04, gao_etal:04,
  de_lucia_etal:04}). These results extend down to subhalos of $M\sim
10^7$ \hmsol\ (\citealt{diemand_etal:07a}).  With the now
well-established connection between satellite galaxies and
substructure within dark matter halos (\citealt{kravtsov_etal:04,
  conroy_etal:06, weinberg_etal:06}), the self-similarity of the HOD
with luminosity is expected to extend far down the mass function to
halos that host dwarf galaxies. In addition, because we are interested
in voids, the details of the $\nsat$ are largely irrelevant; as long
as the fraction of galaxies that are satellites is roughly correct,
the distribution of voids is not affected by the details of $\nsat$
(\citealt{conroy_etal:05, tinker_etal:06_voids}).

For each luminosity threshold, we set the value of $\mmin$ by matching
the space density of galaxies from the \cite{blanton_etal:05}
luminosity function, ie, 

\begin{equation}
\int_{M_r}^\infty \,\Phi(M_r)\,dM_r = \int_0^{\infty}\left(\ncen + \nsat\right)\frac{dn}{dM}\,dM,
\end{equation}

\noindent where $dn/dM$ is the halo mass function, for which we use
\cite{tinker_etal:08}. The \cite{blanton_etal:05} result is accurate
down to $M_r<-12$ galaxies. We set $M_1=22\mmin$ and $\asat=1$ for all
samples. For $\slogm$ we choose a value of 0.15, which is consistent
with the results of \cite{tinker_etal:07_voids} for $M_r<-19$
samples. This is somewhat smaller than that obtained by
\cite{vdb_etal:07}, but we note that the scatter between mass and
luminosity is expected to be larger for blue-selected galaxy samples
(which \citealt{vdb_etal:07} analyze). \cite{tinker_etal:06_voids}
concluded that the values of $\slogm$ ranging from 0 to 0.5 produced
little effect on the void distribution. We will test the dependence of
our results on $\slogm$ in \S 3. Table 1 lists the value of $\mmin$
for each luminosity threshold.

%%%%%%%%%%%%%%%%%%%%%%%%%%%
%        TABLE 2          %
%%%%%%%%%%%%%%%%%%%%%%%%%%%

\begin{deluxetable}{ccccc}
\tablecolumns{5} 
\tablewidth{12pc} 
\tablecaption{Properties of the Simulation Set}
\tablehead{\colhead{$L_{\rm box}$} &  \colhead{$\epsilon$ [\hkpc]} & \colhead{$N_p$} &\colhead{$m_p$ [\hmsol]} &\colhead{$M_r$}}

\startdata

384  & 14 & $1024^3$ &$4.39\times 10^{9}$ & $-20$\\
192  & 4.9 & $1024^3$ &$5.89\times 10^{8}$ & $-17$\\
96 & 1.4 & $1024^3$ &$6.86\times 10^{7}$ & $-14$\\

\enddata \tablecomments{$L_{\rm box}$ is in units of \hmpc. $\epsilon$
  is the Plummer force softening length. $N_p$ is the total number of
  particles in the simulation. Magnitudes in column 5 represent the
  lowest magnitude bin modeled by each simulation.}
\end{deluxetable}

%%%%%%%%%%%%%%%%%%%%%%%%%
% FIGURE
%%%%%%%%%%%%%%%%%%%%%%%%%
\begin{figure*}
\epsscale{0.8}
%\plotone{slice.ps}
\plotone{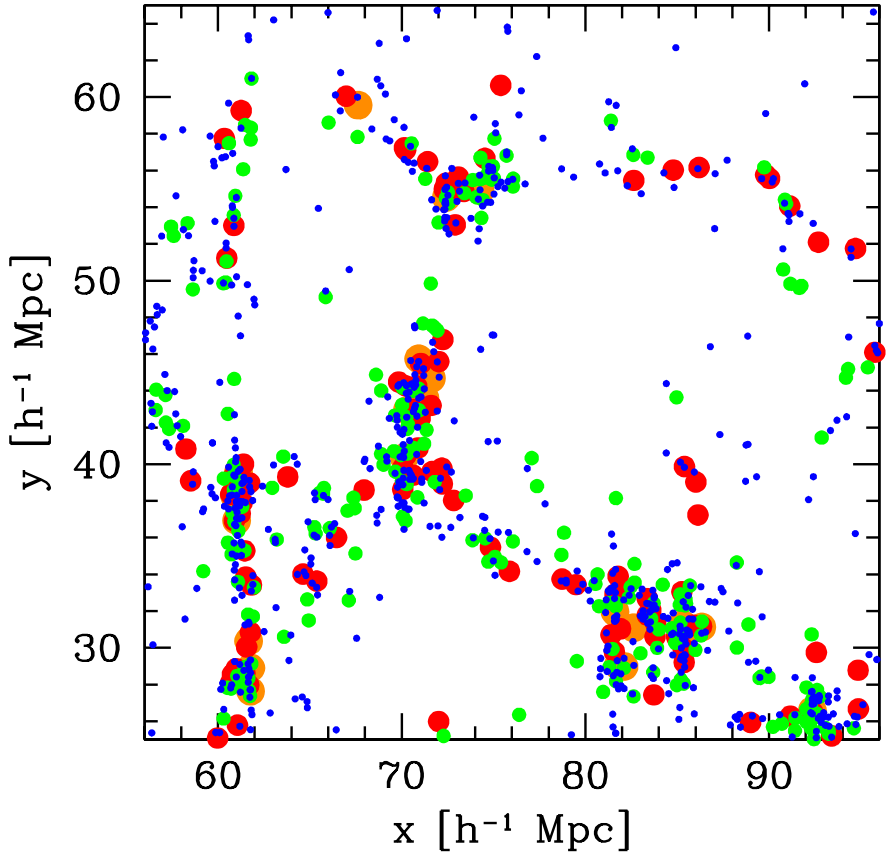}
\caption{ \label{slice} Slice through the galaxy population created
  from the halos within the 96 \hmpc\ simulation. The depth of the
  slice is 6 \hmpc. The point size scales with the luminosity of the
  galaxies. $r$-band magnitudes are as follows: blue $=$ $-14$ and
  $-15$; green $=$ $-16$ and $-17$, red $=$ $-18$ and $-19$, orange are
  all galaxies $-20$ and brighter.}
\end{figure*}
%%%%%%%%%%%%%%%%%%%%%%%%%
% FIGURE
%%%%%%%%%%%%%%%%%%%%%%%%%

%%%%%%%%%%%%%%%%%%%%%%%%%%%%%%%%%%%%%%
\subsection{Numerical Simulations}
%%%%%%%%%%%%%%%%%%%%%%%%%%%%%%%%%%%%%%

To investigate the distribution of voids and void galaxies with our
HOD models, we populate a series of high-resolution N-body simulations
of various box sizes kindly provided my M. Warren. The parameters of
the simulations are listed in Table 2, and we will refer to each
simulation by its box size in \hmpc : L96, L192, and L384. The
simulations allow us to probe halos robustly down to $\sim 10^{10}$
\hmsol, the minimum mass scale of $-12$ galaxies. These simulations
were performed using the hashed oct-tree code of
\cite{warren_salmon:93}, and were described in
\cite{warren_etal:06}. We use the spherical overdensity halo finder of
\cite{tinker_etal:08} to identify halos in the simulation, defining
halos with an overdensity $\Delta=200$ with respect to the background
matter density. The mass resolution of the highest resolution
simulation is 18 times higher than the Millennium Simulation of
\cite{springel_etal:05}, and 75 times higher than the simulation
coupled to the semi-analytic galaxy formation model of
\cite{mathis_white:02}.

To populate each simulation with galaxies, we use a Monte Carlo
approach based on equations (\ref{e.ncen}) and (\ref{e.nsat}),
assuming a nearest integer distribution for central galaxies and a
Poisson distribution of satellite galaxies. The nearest integer
approach is standard for central galaxies due to the limit of one
central galaxy per halo. Poisson statistics provide an excellent
description of the distribution of satellites in simulations
(\citealt{kravtsov_etal:04, zheng_etal:05}) and in observations
(\citealt{lin_etal:04}). By considering the HOD parameters as a
function of luminosity, the full conditional luminosity function of
each halo can be easily determined. The number of centrals in a bin
$M_r$ to $M_r+\Delta M_r$ is $\ncen^{(M_r)}-\ncen^{(M_r+\Delta M_r)}$. The number of
satellite galaxies in a magnitude bin is calculated in the same manner.

Central galaxies are placed at the center of the halo, and satellite
galaxies are distributed randomly with the density profile of
\cite{navarro_etal:97}, with the concentration-mass relation of
\cite{bullock_etal:01} (using the updated parameters of the model
listed in \citealt{wechsler_etal:06}). Satellite velocities, relative
to the halo motion, are taken from a Gaussian distribution with
1-dimensional dispersion of $\sigma^2 = GM_{200}/2R_{200}$. We
populate each simulation down to a limit of 100 particles per
halo. The 384 \hmpc\ simulation is ideal for $M_r<-20$ samples, the
192 \hmpc\ sample resolves $M_r<-17$ samples, while the 96 \hmpc\
simulation probes galaxies as faint as $M_r=-14$. Although $\mmin$ for
$M_r=-12$ is 127 particles, we must resolve halos below $\mmin$ to
fully account for the scatter between mass and luminosity.

\section{Results}

%%%%%%%%%%%%%%%%%%%%%%%%%
% FIGURE
%%%%%%%%%%%%%%%%%%%%%%%%%
\begin{figure}
\epsscale{1.17}
%\plotone{void_LF.ps}
\plotone{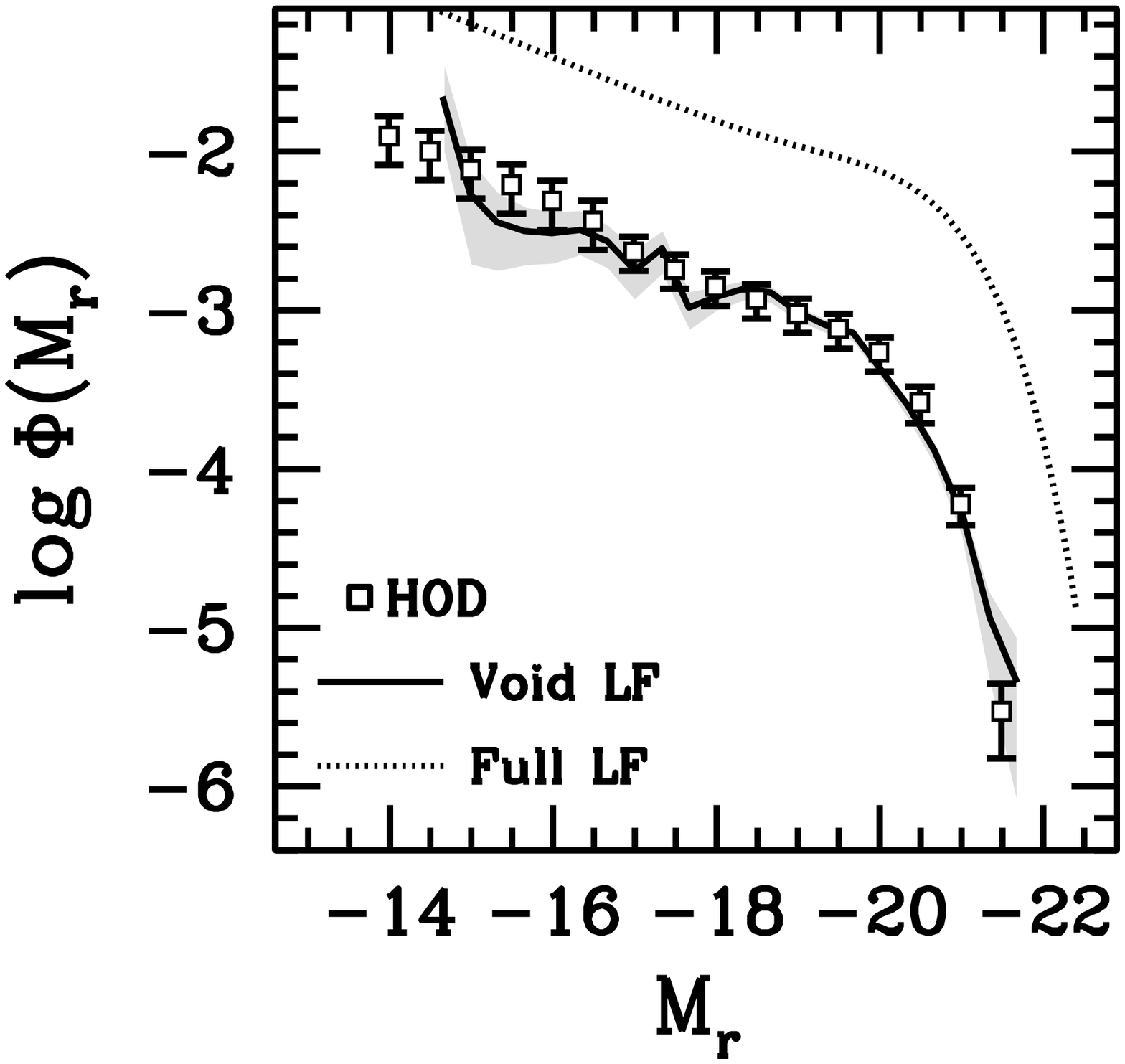}
\caption{ \label{void_LF} Comparison between the luminosity function
  of void galaxies measured by \cite{hoyle_etal:05} and predicted by
  our HOD model. The HOD measurement, shown with the open squares, is
  a weighted average of results from the 96 \hmpc\ and 192 \hmpc\
  simulations (noting that all points at $M_r>-17$ are from the 96
  \hmpc\ simulation). Errors are estimated by jackknife sampling of
  each simulation into octants. The solid line is the Hoyle et al.\
  data, with the shaded region representing their error bars. For
  comparison, the full luminosity function of all galaxies from
  \cite{blanton_etal:05} is shown with the dotted curve.}
\end{figure}

\subsection{The Cosmic Web in the Faintest Galaxies}

Figure \ref{slice} presents a slice through L96 that is typical of the
structure in these simulations. The galaxy positions are in redshift
space, assuming the distant-observer approximation with the $y$-axis
being the line of sight. Along the filaments, $L_\ast$ and dwarf
galaxies cohabitate, along with all luminosities in between. The
transition between filaments and voids is very sharp, however. Deep
voids of $\sim 10$ \hmpc\ in diameter, empty of even the
faintest resolved galaxies, are not uncommon in this simulation. The
deepest void, centered at $(x,y)=(80,48)$ \hmpc, is nearly $18$ \hmpc\
along its longest axis. The transition from the main filament in the
slice (running along the $y$-axis) into the void on either side is
sharp for both bright and faint galaxies, resembling the galaxy
distribution around the local void shown in \cite{peebles:07}. Around
the local void, between 5 to 7 \hmpc\ from the void center, the
number of galaxies $M_B<-18$ and $M_B>-18$ drops essentially to
zero. The abrupt transition from filament to void is ubiquitous in the
cosmic web, even for galaxies hosted by $10^{10}$ \hmsol\ halos. We
note that halo velocities are usually directed along the axis of the
filament, thus choosing the $x$-axis as the line of sight does not
noticeable change the galaxy distribution plotted in Figure
\ref{slice}.

\subsection{The Luminosity Function of Void Galaxies}

The most straightforward quantifiable test of our model is the number
of galaxies located within voids, and their distribution as a function
of luminosity. \cite{hoyle_etal:05} created a sample of ``void''
galaxies in the SDSS by identifying galaxies with local densities
below a critical value of $\delta\equiv\delta\rho/\rhob= -0.6$ in a
top-hat sphere of 7 \hmpc. Densities were calculated with respect to
galaxies with $M_r<-20.5$. We have constructed a sample of void
galaxies in our simulations following these same criteria. All
densities are calculated in redshift space assuming the distant
observer approximation. Figure \ref{void_LF} compares the void
luminosity function of our HOD model to the \cite{hoyle_etal:05}
measurements. Our model is in excellent agreement with the data, not
only reproducing the overall abundance of void galaxies but also the
decrease in the value of $M_\ast$ in the void luminosity function
relative to the overall luminosity function. 

Interestingly, we find some void galaxies with magnitudes as bright as
$M_r=-21.5$, just as measured in Hoyle et al. These objects are {\it
  not} scattered into void regions due to redshift space distortions,
but are in intrinsically low-density regions as defined by the
$M_r<-20.5$ galaxies. The minimum mass scale for $M_r<-21.5$ galaxies
is $\sim 1.8 \times 10^{13}$ \hmsol, which are never found in
$\delta\lesssim -0.6$ regions in the {\it dark matter}
distribution. When using the dark matter particles to obtain the local
density around each galaxy and recalculate the void luminosity
function, there are no void galaxies this bright. Thus, bright void
galaxies are due to stochastic biasing of $L_\ast$ galaxies, both from
Poisson fluctuations in the number of halos and fluctuations in the
number of $L>\lstar$ galaxies per halo. This creates a few regions in
which the dark matter density is above the density threshold but the
galaxy density is below it.

%%%%%%%%%%%%%%%%%%%%%%%%%
% FIGURE
%%%%%%%%%%%%%%%%%%%%%%%%%
\begin{figure*}
\epsscale{1.}
\vspace{-1.1cm}
%\plotone{nnb9.ps}
\plotone{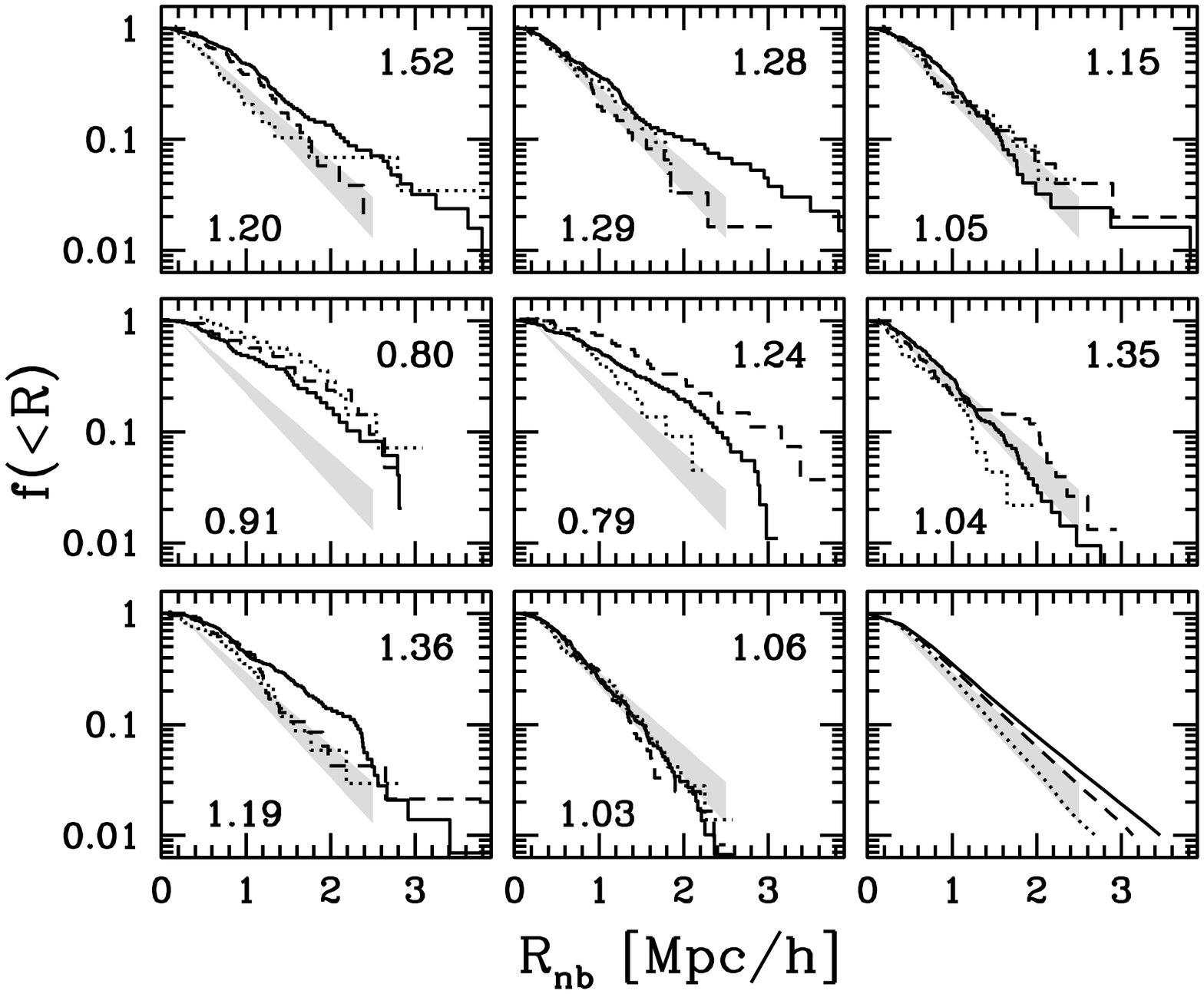}
\caption{ \label{nnb9} Nearest neighbor statistics for ``test''
  galaxies, $M_r=[-14,-16)$ (solid histograms), ``control'' galaxies,
  $M_r=[-16,-18)$, (dashed histograms), and ``bright'' galaxies,
  $M_r<-18$, (dotted histograms). For each galaxy in a sample, $\rnb$
  is the distance to the nearest control galaxy in redshift space. The
  first eight panels represent random subvolumes of the 96 \hmpc\
  simulation that match the volume of the sample in P01. In each
  panel, the number in the bottom left is the ratio of mean $\rnb$
  values, $\langle R_{tc}\rangle/\langle R_{cc}\rangle$. The number in
  the top right is $\langle R_{tc}\rangle/\langle R_{bc}\rangle$. The
  bottom right panel are the overall statistics for the entire
  box. The shaded region represents the observational results from
  P01.}
\end{figure*}
%%%%%%%%%%%%%%%%%%%%%%%%%
% FIGURE
%%%%%%%%%%%%%%%%%%%%%%%%%

\subsection{Nearest Neighbor Statistics}

P01 used nearest neighbor statistics to probe the relative
distribution of ``ordinary'' galaxies ($M_r\le -16$) to dwarf galaxies
($M_r>-16$). If dwarf galaxies preferentially occupy void regions
relative to their brighter counterparts, their distribution of nearest
neighbors will show a significant tail out to large neighbor distances
$\rnb$. To circumvent the problem of different galaxy samples having
different mean space densities, for each test object the nearest
neighbor in the control sample is found. This distribution is compared
to the distribution of $\rnb$ of the control sample to itself. P01
found that the distribution of nearest neighbors for test and control
objects are essentially the same, indicating that dwarfs avoid the
voids defined by the ordinary objects.

In Figure \ref{nnb9} we present several examples of the cumulative
distributions of $\rnb$ taken from L96. Here we have set our control
sample to be $M_r=[-16,-18)$, and our test sample are galaxies with
$M_r=[-14,-16)$. The control and test samples are substantially
different in terms of their space densities, but are not too
dissimilar in the halo masses that they probe. To test any systematics
due to these choices we have an additional sample of ``bright''
objects containing all galaxies $M_r<-18$. The largest sample in P01
is roughly 1000 $($\hmpc$)^3$, so we have broken our mock galaxy
distribution into $9^3$ equal-volume cubes of 1225 $($\hmpc$)^3$ to
test for cosmic variance in this statistic. The bottom right panel
shows the results from the full simulation. We define $R_{tc}$ as the
distance to the nearest control galaxy for each test galaxy, while
$R_{bc}$ is the distance to the nearest control galaxy for each bright
galaxy, and $R_{cc}$ is the distance for each control galaxy to the
nearest galaxy within the control sample.

The shaded region in each panel of Figure \ref{nnb9} approximates the
results from the largest dwarf sample from P01 (his Figure 4, top set
of curves). The upper limit of the shaded region is set by
$f(<R_{tc})$ while the lower edge is $f(<R_{cc})$. P01 determined the
ratio of mean $\rnb$ values to be $\langle R_{tc}\rangle/\langle
R_{cc}\rangle=1.1$. This is nearly identical to the overall value of
1.06 obtained from the L96 simulation. The scatter in this statistic
is quite large, with a variance in mean distance ratios over the $9^3$
subsamples of 0.23. In many examples in Figure \ref{nnb9},
$f(<R_{cc})$ extends to distances significantly beyond
$f(<R_{tc})$. Comparing the $R_{tc}$ distributions to the nearest
neighbor statistics for the bright objects to the test sample,
$R_{bc}$, yields similar results. The mean distance ratio is 1.19 with
a scatter of 0.55, owing to the lower number density of brighter
objects. While the agreement between model and data are encouraging,
it is likely fortuitous given the large cosmic scatter in the P01
measurements. However, we conclude that the prediction of the model
supports the picture that dwarf galaxies and brighter galaxies have
similar spatial distributions in low-density regions. The slight
difference in nearest neighbor distributions between dwarf and regular
galaxies is also in agreement with the observational results of
\cite{lee_etal:00}.

%%%%%%%%%%%%%%%%%%%%%%%%%
% FIGURE
%%%%%%%%%%%%%%%%%%%%%%%%%
\begin{figure}
\epsscale{1.2}
%\plotone{void_sizes.ps}
\plotone{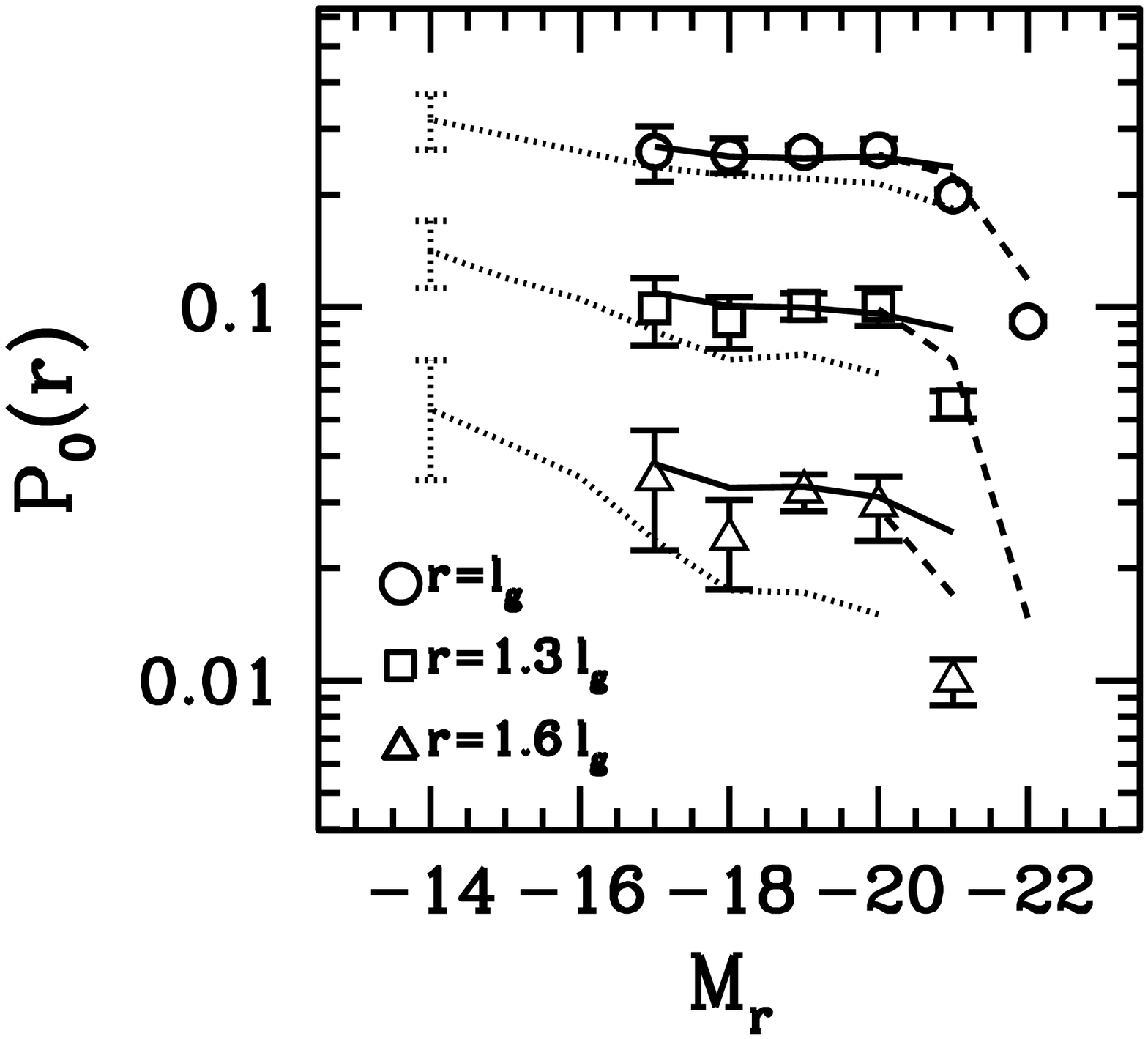}
\caption{ \label{void_sizes} The void probability as a function of
  luminosity at three different separations relative to the mean
  intergalactic separation, $\lg=\ngavg^{-1/3}$. Points with errors
  are results from SDSS DR6. Curves represent results from the HOD
  model from the simulations listed in Table 2: L96 ({\it dotted
    curves}), L192 ({\it solid curves}), L384 ({\it dashed
    curves}). Data points for the $M_r=-22$ sample are not shown for
  $r\ge 1.3\lg$ because the VPF is not measured at those distances.}
\end{figure}
%%%%%%%%%%%%%%%%%%%%%%%%%
% FIGURE
%%%%%%%%%%%%%%%%%%%%%%%%%

\subsection{Void Probabilities as a Function of Luminosity}

The void probability function (VPF, denoted $\p0$) is defined as the
probability that a randomly placed sphere of radius $r$ contains no
galaxies. In \cite{tinker_etal:07_voids} we measured the VPF for
galaxies as faint as $M_r=-19$ from Data Release Four of the Sloan
Digital Sky Survey (\citealt{dr4}).  Fainter samples were not used
because cosmic variance errors become large and galaxy clustering
measurements are not available at $M_r>-18$. In Figure
\ref{void_sizes} we present new measurements of the VPF from DR6. The
measurements are in magnitude bins, 1 magnitude wide (referenced by
their lower limit). We follow the procedures outlined in detail in
\cite{tinker_etal:07_voids} for both measurements and for comparing
the HOD predictions to the data. We use the HOD mocks to estimate
error bars on the data, also discussed in
\cite{tinker_etal:07_voids}. The increased sky coverage of DR6
attenuates (but does not eliminate) the cosmic variance considerations
for fainter samples. Though we are not able to probe void statistics
for dwarf galaxies (in our definition), we are able to make VPF
measurements down to $M_r=-17$, a two magnitude improvement on our
previous results. We use these data to test the robustness of our HOD
model and to extend the conclusions of \cite{tinker_etal:07_voids}
discussed in \S 1 to lower luminosities.

Rather than presenting VPFs for each luminosity bin, we consolidate
the results by presenting void probabilities at a fixed distance scale
as a function of luminosity. The open circles in Figure
\ref{void_sizes} show $\p0$ at the mean intergalactic separation for
each magnitude bin, $r=\lg\equiv\ngavg^{-1/3}$. Squares and triangles
represent $\p0$ at $r=1.3\lg$ and $r=1.6\lg$, respectively. At
luminosities below $\lstar$, the void probability at any multiple of
$\lg$ is essentially independent of luminosity. Brighter galaxies have
a lower probability of finding a void, thus, in the scaled distance
$r/\lg$, the brightest galaxies have the smallest voids. This has also
been seen in the 2dFGRS data (\citealt{beckmann_muller:08}). This is
partially a result of the fact that brighter galaxies have fewer
satellite galaxies (a higher satellite fraction results in an increase
in $\mmin$ in order to match the number density of the galaxy sample),
but it is mainly due to the fact that $\lg$ for samples on the
exponential tail of the of the luminosity function increases much more
rapidly than the bias of those galaxies. For $M_r<-22$ galaxies,
$\lg=33$ \hmpc.

The HOD model predictions are shown in Figure \ref{void_sizes} for all
three simulations. The difference in the amplitude of $\p0$ between
L96 and the larger simulations is consistent with cosmic variance due
to the small volume. Although the model is not calibrated on
clustering measurements as done in \cite{tinker_etal:07_voids}, it
matches the observed void probabilities as a function of luminosity
and scale. The constraints on $\mmin$ are driven primarily by $\ngavg$
rather than two-point clustering, therefore $\mmin$ and the resulting VPF
are similar between \cite{tinker_etal:07_voids} and the model
presented here.

Although we cannot probe the VPF for galaxies as faint as $M_r=-14$
the success of our model in matching $\p0$ down to $M_r=-17$, the void
luminosity function down to $M_r=-14.5$, and nearest neighbor
statistics down to $M_r=-14$ strongly suggest that our extrapolation of
the HOD below halo masses of $10^{11}$ \hmsol\ is robust, and the
resulting galaxy distribution represents a complete picture of the
structure within underdense regions.

\subsection{ The galaxy structure within voids}

In the VPF results of Figure \ref{void_sizes}, $P_0(1.6\lg) \sim
{constant}$ for $-17\ge M_r\ge -20$ galaxies in both the model and the
data. In this magnitude range, voids are self-similar. At fainter
magnitudes, the void probability monotonically increases with
decreasing brightness. This implies that voids are not self-similar
for dwarf galaxies, and that the structure of the comic web itself
plays a role in the distribution of dwarfs in voids (as seen in Figure
\ref{slice}).

\begin{figure}
\epsscale{1.2}
%\plotone{void_profile.ps}
\plotone{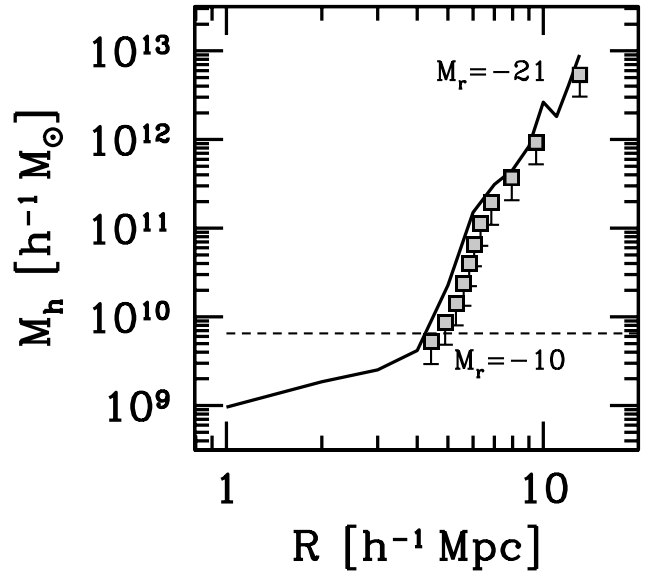}
\caption{ \label{void_profile} The maximum halo mass within voids as a
  function of distance from the void center. The solid curve
  represents mean of the largest three voids in the 96 \hmpc\
  simulation. The points represent $\mmin$ as a function of magnitude,
  placed along the curve to demonstrate how large the void would be at
  each luminosity. The dashed curve is the critical mass from
  \cite{hoeft_etal:06}, below which star formation in void halos is
  significantly attenuated due to delayed formation histories of void
  halos.}
\end{figure}

The results of Figure \ref{slice} and the trend in $\p0$ for $M_r\ge
-17$ galaxies can be interpreted through the change in the halo mass
function within the void itself. Figure \ref{void_profile} plots the
{\it maximum} halo mass as a function of distance from the center of
the three deepest voids in L96 (all with radii $\sim 10$ \hmpc, and
central densities $\delta<-0.9$). At 10 \hmpc\ from the center of the
void, halos between $10^{12}$-$10^{13}$ \hmsol\ can be found. These
halos house the $\lstar$ galaxies that `define' the edge of the
void. But between 10 \hmpc\ and 5 \hmpc\ from the void center, the
maximum halo mass drops nearly three orders of magnitude. At the very
centers of the voids, only $M\sim 10^9$ \hmsol\ halos are present. The
points in Figure \ref{void_profile} show $\mmin$ as a function of
luminosity from $M_r<-21$ to $M_r<-10$ samples\footnote{We caution
  that the \cite{blanton_etal:05} luminosity function is calibrated
  only down to $-12$, so results at lower luminosities are an
  extrapolation.}. The points are placed along the curve to
demonstrate how large the void would be in each luminosity. The steep
drop in the maximum mass at $R \sim 6$ \hmpc\ causes the voids to have
sharp boundaries in the galaxy distribution over a wide range of
galaxy luminosity. Figure \ref{void_profile} implies that between 5
and 6 \hmpc\ from the void center, the brightest galaxy luminosity
will plummet by nearly 5 magnitudes. The galaxies in the inner half of
the void (if there are any) would be well below the magnitude limit of
the ORS data with which P01 calculate their
statistics. \cite{peebles:07} finds that the abrupt transition into
the local void, inside of which no galaxies are observed, occurs at
$R\sim 5-7$ \hmpc\ from the void center. He further demonstrates that
this boundary is nearly identical for galaxies divided into faint and
bright samples (with a threshold of $M_B=-18$).  The scatter between
halo mass and galaxy luminosity smooths out the slight change in void
radius with $M_r$ seen in Figure \ref{void_profile}. Thus for a
luminosity threshold cut, the edge of the local void should be nearly
the same for the two samples.

The dashed curve in Figure \ref{void_profile} indicates the critical
mass threshold below which \cite{hoeft_etal:06} find that
photoionization significantly reduces the ability of void halos to
form stars. No halos at $R\le 5$ \hmpc\ are above this line,
suggesting that the inner half of deep voids will contain no galaxies
at all, becoming truly dark. \cite{gottlober_etal:03} calculated the
mean void mass function for all halos within 80\% of the void radius,
finding good agreement with the analytic predictions of
\cite{sheth_tormen:02}. However, they also find that the shape of the
void mass function depends on position within the void itself. The
{\it mean} void mass function in the L96 simulation (within $0.8\rv$)
is in good agreement with their results.

It is important to consider scatter between mass and luminosity for
central galaxies. The points in Figure \ref{void_profile} represent
$\mmin$, the mass at which the halo has only a 50\% probability of
containing a galaxy of that luminosity or brighter. The error bars on
each point indicate the mass at which that probability is $10\%$ {\it
  and} if we {\it double} the scatter between mass and luminosity to
0.3. Random fluctuations, or larger physical scatter between mass and
luminosity, may place a galaxy slightly further toward the center of
the void (or further away), but not enough to smooth out the sharp
transition between filament and void seen in the simulations.

%%%%%%%%%%%%%%%%%%%%%%%%%
% FIGURE
%%%%%%%%%%%%%%%%%%%%%%%%%
%%%%%%%%%%%%%%%%%%%%%%%%%
% FIGURE
%%%%%%%%%%%%%%%%%%%%%%%%%

%%%%%%%%%%%%%%%%%%%%%%%%%
\section{Summary and Discussion}
%%%%%%%%%%%%%%%%%%%%%%%%%

The void phenomenon consists of two observational facts: that voids
contain few, if any, low-luminosity galaxies, and that the few void
objects tend to have similar properties to the overall galaxy
population. The controversial aspect is whether these facts are at
odds with the current cosmology. Although the depth of voids and
homogeneity of void objects are striking features of the cosmic web,
they are readily explainable within the context of \lcdm, combined
with a straightforward model to connect galaxies and dark matter at
all luminosities and mass scales.

The simple proposition within our implementation of the halo
occupation distribution is that galaxy properties are determined
solely by the mass of the halo in which the galaxy resides,
independent of the halo's larger scale environment\footnote{We note
  that it is possible to create an HOD with depends on
  environment. The model would then specify $P(N|M,\delta)$ rather
  than $P(N|M)$, as done in \cite{tinker_etal:06_voids,
    tinker_etal:07_voids} and \cite{wechsler_etal:06}.}  Although this
model must break down at some high level of precision or with the
details of certain galaxy properties, connecting luminosity to halo
mass is a robust method that has passed all tests thus far
(\citealt{abbas_sheth:06, skibba_etal:06, tinker_etal:07_voids}, as
well as the tests presented in this paper). The mass-only approach to
galaxy bias readily explains both aspects of the void
phenomenon. First, void galaxies are a fair representation of the
field population simply by construction. A galaxy in a $10^{10}$
\hmsol\ halo does not know if the halo sits in a void or a
filament. Recent numerical results have demonstrated that the halo
itself does retain information about its environment (the 'assembly
bias'; see, e.g., \citealt{gao_etal:05, wechsler_etal:06,
  gao_white:07}). However, this effect does not propagate into the
void galaxies through either luminosity or color
(\citealt{tinker_etal:07_voids}). \cite{patiri_etal:06b} find that
void galaxy properties, even star formation rate or morphology, follow
the same distribution as the field. This is echoed in the
semi-analytic results of \cite{croton_farrar:08}. Therefore, although
halos 'remember' their environment, the physics of galaxy formation
introduces an intrinsic scatter between halo formation and galaxy
properties that washes this signal out.

Second, the lack of dwarf galaxies in voids is a consequence of
mapping galaxy luminosity to halo mass in such a way as to preserve
the observed abundance of galaxies at each luminosity. The halo mass
function scales as $M^{-2}$, while the luminosity function scales as
roughly $L^{-1}$. This difference in logarithmic slopes implies that
the mass to light ratio $M/L$ of galaxies increases as galaxies become
fainter. Below $\lstar$, objects with substantially different
luminosities will occupy halos of roughly comparable mass and will
trace out the same large-scale structure. In order to explain the void
phenomenon, we require the physical mechanism that causes the increase
in $M/L$ for faint objects (whatever it may be) to apply to all
low-mass halos, regardless of environment.

Connecting the luminosity function to the halo mass function naturally
predicts that the Tully-Fisher relation becomes flatter at low
luminosities. \cite{geha_etal:06} confirmed this prediction by
measuring the rotation velocities of galaxies as faint as $M_r\approx
-13$. Our model places these galaxies in halos with maximum circular
velocities of $\sim 45$ \kms, in good agreement with the observed
rotation speeds of 35-60 \kms. 

As stated above, parameterizing the HOD as a function only of halo
mass is only an approximation of the underlying physics (although a
surprisingly good approximation). In the context of the void
phenomenon, it should be noted that any influence of environment is
mostly likely to {\it reduce} galaxy formation efficiency in
underdense regions. This is the effect of the photoionization
arguments in \cite{hoeft_etal:06}, and is also seen in the
semi-analytic model of \cite{croton_etal:07}. A reduction in formation
efficiency in voids makes $\mmin$ {\it higher} at fixed
luminosity. There is no need to invoke exotic new physics, such as
modified gravity models, to lower the formation efficiency of halos in
underdense regions to match the data. Several studies have
demonstrated that a long-range scalar potential acting on the dark
matter can accelerate the evacuation of matter from voids during the
growth of structure (\citealt{farrar_peebles:04, gubser_peebles:04,
  nusser_etal:05}). Although comparison of these models to
observational data is difficult analytically, numerical simulations of
these alternative gravity models, like those by \cite{nusser_etal:05},
can be used in the same manner as the simulations here. The excellent
agreement between our HOD model and the observations may place upper
limits on such models, since models that produce larger voids at any
luminosity would come into conflict with measured void statistics.

\vspace{1cm}

\noindent We would like to thank Mike Warren for use of his
simulations. We also thank Jim Peebles for comments on the manuscript.
J.T. was supported by the Chandra award GO5-6120B and National Science
Foundation (NSF) under grant AST-0239759.

%%%%%%%%%%%%%%%%%%%%%%%%%%%%%%%%%%%%%%%%%%%%%%%%%%%%%%%%%%%%%%%%%%%%%%%%
%  Bibliography
%%%%%%%%%%%%%%%%%%%%%%%%%%%%%%%%%%%%%%%%%%%%%%%%%%%%%%%%%%%%%%%%%%%%%%%%

\bibliography{../risa}

\end{document}